\documentclass[a4paper,%
 aip,apm
 sd,%
 amsmath,amssymb,
preprint,%
]{revtex4-1}

\usepackage{graphicx}
\usepackage{dcolumn}
\usepackage{bm}
\usepackage{color}

\usepackage{xspace}
\usepackage{url}
\newcommand{\sto}{$\mathrm{SrTiO}_3$\xspace}

\newcommand{\degree}{\ensuremath{^\circ}}
\newcommand{\cel}{\degree C}
\newcommand{\tgd}{\tan \delta}
\newcommand{\epsr}{\epsilon_r'}
\newcommand{\epsf}{\epsilon_{eff}}
\newcommand{\epsrom}{\epsilon_r(\omega)}
\usepackage[breaklinks=true]{hyperref}
\usepackage{color}

\begin{document}

\preprint{AIP/123-QED}

\title[High permittivity processed \sto for metamaterials applications at  terahertz]{High permittivity processed \sto for metamaterials applications at  terahertz frequencies}

\author{Cyrielle Dupas}
\affiliation{Univ Limoges, CNRS, SPCTS UMR 7315, Ctr Europeen Ceram, F-87068 Limoges, France}
\affiliation{Univ. Paul Sabatier, CNRS, Institut Carnot, CIRIMAT, UMR 5085, F-31062 Toulouse, France}
\author{Sophie Guillemet-Fritsch}
\affiliation{Univ. Paul Sabatier, CNRS, Institut Carnot, CIRIMAT, UMR 5085, F-31062 Toulouse, France}
\author{Pierre-Marie Geffroy} 
\affiliation{Univ Limoges, CNRS, SPCTS UMR 7315, Ctr Europeen Ceram, F-87068 Limoges, France}
\author{Thierry Chartier}
\affiliation{Univ Limoges, CNRS, SPCTS UMR 7315, Ctr Europeen Ceram, F-87068 Limoges, France}
\author{Matthieu Baillergeau}
\affiliation{Univ Paris 06, Univ D. Diderot, CNRS, Ecole Normale Super, Lab Pierre Aigrain,UMR 8551, F-75231 Paris 05, France}
\author{Juliette Mangeney}
\affiliation{Univ Paris 06, Univ D. Diderot, CNRS, Ecole Normale Super, Lab Pierre Aigrain,UMR 8551, F-75231 Paris 05, France}
\author{Jean-Pierre Ganne}
\affiliation{Thales Research \& Technology, Route D\'epartementale 128, 91767 Palaiseau Cedex, France}
\author{Simon Marcellin}
\affiliation{Institut d'\'Electronique Fondamentale, Univ. Paris-Sud, Universit\'e Paris-Saclay, Orsay, F-91405; UMR8622, CNRS, Orsay, F 91405.}
\author{Aloyse Degiron} 
\affiliation{Institut d'\'Electronique Fondamentale, Univ. Paris-Sud, Universit\'e Paris-Saclay, Orsay, F-91405; UMR8622, CNRS, Orsay, F 91405.}
\author{\'Eric Akmansoy}
\email{{eric.akmansoy@u-psud.fr}}
\affiliation{Institut d'\'Electronique Fondamentale, Univ. Paris-Sud, Universit\'e Paris-Saclay, Orsay, F-91405; UMR8622, CNRS, Orsay, F 91405.}

\date{\today}

\begin{abstract}
High permittivity \sto for the realization of all-dielectric metamaterials operating at terahertz frequencies was fabricated. A comparison of different processing routes evidences that Spark Plasma Sintering is the most effective sintering process to yield high density ceramic with high permittivity. We compare this sintering process with two others. The elaborated samples are characterized in the low frequency and in the terahertz frequency ranges. Their relative permittivities are compared with that a reference \sto single crystal. The permittivity of the sample elaborated by Spark Plasma Sintering is as high as that the single crystal. 
\end{abstract}

\pacs{78.20.Ci, 77.22.Ch, 81.05.Mh, 78.67.Pt.}
\keywords{Dielectrics, Ceramics, Spark Plasma Sintering, THz Time Domain Spectroscopy, Metamaterials}
\maketitle

All-Dielectric Metamaterials (ADM) are the promising alternative to Metallic Metamaterials (MM). Metamaterials give rise to unnatural phenomena such as negative index, sub-wavelength focusing and cloaking. They are engineered materials whose unit cell generally comprises two sub-wavelength building blocks. 
MMs were firstly demonstrated in the microwave. However, going up to the terahertz and the optical domains has been difficult due to ohmic losses and complicated geometries. ADMs rely on the first two modes of Mie resonances of High Permittivity Resonators (HPR).\cite{prb77_mosallaei, prl100_popa} They do not suffer from ohmic losses and consequently benefit of low energy dissipation;\cite{prl100_popa} moreover, their unit cell is of simple geometry. HPRs at a few tens of microns scale are required for ADMs applications in the THz range.\cite{jstqe_gaufillet} 

Terahertz (THz) radiation is widely defined as the electromagnetic radiation in the frequency range 0.3-3 THz. It permits to obtain physical data which are not accessible by the means of X-rays or infrared radiation. In this respect, THz radiation offer many applications in imaging, spectroscopy, chemical sensing, astronomy, security, etc. On their part, metamaterials have evolved towards the implementation of optical components.\cite{nm11_kivshar}
ADMs permit to achieve a great number of fascinating phenomena (see ref.\,\onlinecite{nn1_jacob} for a review). 
They nevertheless require efficient fabrication processes to develop, namely,  processes which lead to high permittivity ceramics that could be structured at the micron scale. Herein, we show that Spark Plasma Sintering makes it possible to fabricate dense high permittivity \sto ceramic suitable for applications at terahertz frequencies. We compare this sintering process with two others routes: tape casting and uniaxial pressing. 
The structural properties of the samples were investigated by X-ray diffraction (XRD) and Scanning Electron Microscope (SEM). 
Then, the samples were characterized in the low frequency range and at THz frequencies by the means of Time Domain Spectroscopy (THz--TDS), and their dielectric constant was compared with that of a reference \sto single crystal.


Commercial \sto powder \footnote{Marion Technologies \url{http://www.mariontechnologies.com/nanomateriaux/}}, with a mean grain size of 0.5$\mu$m, is used to shape ceramic according to three different processes: Tape Casting (TC), Uniaxial Pressing (UP) and Spark Plasmas Sintering (SPS), the two former involving conventional sintering. Each fabrication process was implemented to prepare a batch of a few samples under similar conditions from the same \sto powder. Their chemical composition is consequently the same.


Tape casting makes it possible to fabricate thin ceramic sheets with controlled thickness ranges from several tens to a few hundred micrometers.\cite{chartier}
%
A suspension consisting of the ceramic powder dispersed in a solvent, with the help of a dispersant and containing a binder and a plasticizer,  is cast onto a fixed support (Mylar\textsuperscript{\textregistered} film). Once the solvent is evaporated, the thickness of the flexible green ceramic tape is around 100\,$\mu$m. Disks are cut in the tape by a laser beam to avoid stresses in the green tape. These disks are then debinded and pre-sintered at 1100\cel, before undergoing conventional sintering at a temperature between 1320\cel{ } and 1350\cel{Ê} in air during 1h. These samples are referred to as TC hereinafter. 

The \sto powder is dispersed in water with addition of a binder, then granulated by spray-drying. A controlled amount of \sto granules is uniaxially pressed into a steel matrix at a pressure of 200 MPa. Obtained green ceramic pellets are then debinded and pre-sintered at 1100\cel{}, before conventional sintering at 1330\cel{}. Then, the ceramics pellets are polished until their thickness is a few hundreds micrometers. 
%
These samples are quoted UP hereinafter.

Spark Plasma Sintering is a sintering process which relies on the heating by a pulsed electric current combined with high uniaxial pressure (around 75MPa).\cite {msea287_omori, sss5_nygren}  The \sto powder is set into a carbon graphite dies, through which the current is conducted. This process has several advantages because it allows fast heating and the possibility to obtain fully dense samples at comparatively lower sintering temperatures. The grain growth is greatly reduced, while the ceramic samples rapidly get very dense (usually above 98\% in a 20 min cycle) (\textcolor{blue}{See supplementary material}). 
The sintered \sto ceramics are dark blue colored, which typifies the presence of Ti$^{3+}$ cations. \cite{IJAC10_guillemet} Indeed, during the sintering, the reducing atmosphere reduces Ti$^{4+}$ cations into Ti$^{3+}$ cations. Samples are then annealed at 850\cel{ }during a couple of hours  in air so as to re-oxidize the Ti$^{3+}$ cations in the sample. These samples are a few hundred micrometers thick and are  quoted SPS hereinafter.

\begin{figure}[htbp]
\begin{center}
\includegraphics[width =\linewidth]{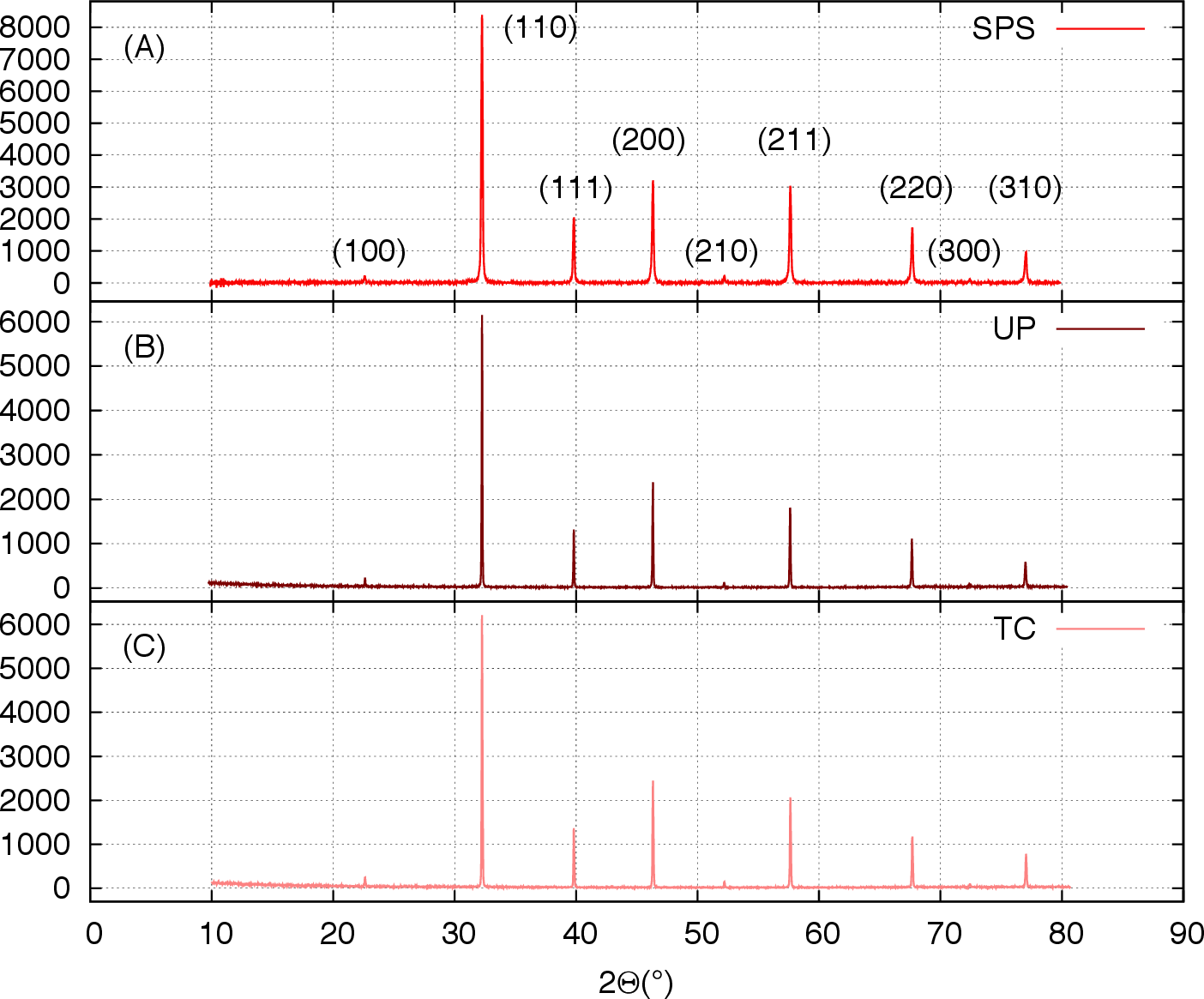}
\caption{: XRD patterns of sintered \sto samples: (a) SPS sample, (b) UP sample, (c) TC sample. The samples are in the same crystalline phase (JCPDS 01-070-8508).}
\label{fig:xrd}
\end{center}
\end{figure}

The crystalline structure and phase purity were firstly observed at room temperature \textit{via} XRD measurements using a Bruker D4 diffractometer. 
Then, the grain size and the morphology of the sintered ceramics were observed using SEM. 
Further, dielectric characterization was carried out in the kHz and the THz ranges in order to determine the dielectric constant (relative permittivity $\epsr$ and dielectric losses $\tgd$). 
On the one hand, the low frequencies dielectric measurements (100\,Hz -- 1 MHz) were performed by the means of an Impedance Analyzer Agilent 4294A. 
In the kHz range, the dielectric constant mainly depends on the chemical composition, the structure,  the grain size and the density of the ceramics.\cite{mcp83_li}
On the other hand, the terahertz dielectric properties were measured by the means of a THz--TDS set up. \cite{sr6_mangeney} 
The TDS setup is based on a Ti:Sa laser (pulse duration 15 fs, central wavelength 800 nm) that is split into a pump and a probe beams. The former is converted into THz radiation (pulse duration in the picosecond range) using a photoconducting antenna patterned on a low-temperature GaAs substrate. It is then focused on the sample and the transmitted signal is measured \textit{via} electro-optic detection. To this aim, it is combined with the probe beam which samples the THz signal thanks to a delay line.
Using this technique, the temporal shape of the THz signal and its delay with respect to a reference measurement, made without sample, are obtained. The data are subsequently Fourier transformed to obtain a transmission spectrum in the frequency domain. Finally, the values of the optical index is derived from these measurements by inverting the Fresnel equations that describe the THz transmission through the samples. \cite{jstqe2_duvillaret, mtt53_kurz}



\begin{figure}[!htp]
\includegraphics[width =\linewidth]{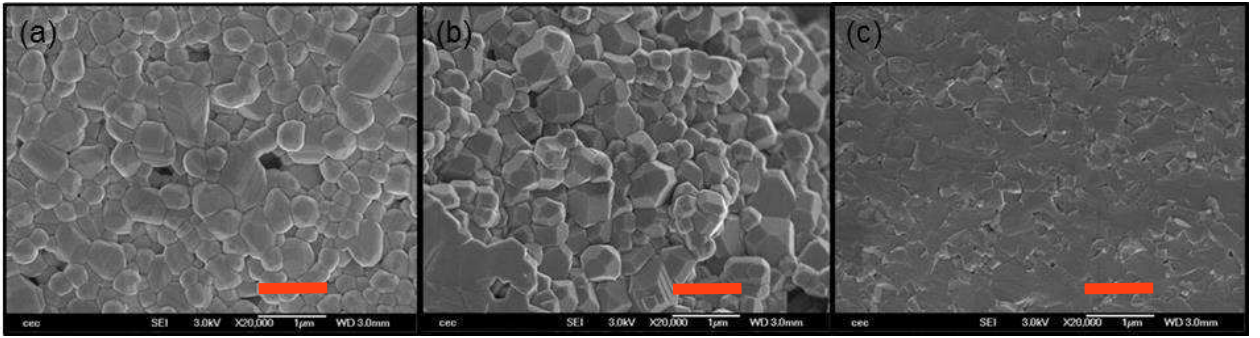}
\caption{\label{fig:sem} SEM images of sintered \sto samples: (a) TC sample, (b) UP sample, (c) SPS sample. Scale = 1$\mu$m (orange dash). Porosity can be noticed in the TC sample.} 
\end{figure}

XRD analyses were performed on the batches of sintered ceramics (Fig.\,\ref{fig:xrd}). Whatever the shaping process, all the samples are in the same crystalline phase, namely, the cubic perovskite structure of \sto. 
Then, the relative density $d$ of the samples was measured by Archimed's method (Table\,\ref{tab:table1}).\cite{jjap33_sasaki}  The relative density of the TC samples is low: $d\!\approx\!70\%$, whereas that of the UP samples is improved, ranging from $d\!=\!90$\% to 95\%.
At last, the SPS samples exhibit the highest density: $d\!\geq\!99\%$. 
In addition, whatever the shaping process, the mean grain size of the samples, observed by SEM, is about 0.5$\mu$m (Fig.\,\ref{fig:sem}). Consequently, as the chemical composition, the structure and the grain size are identical whatever the implemented process, only the density of the samples may affect their dielectric constant. 

\begin{table}
\caption{\label{tab:table1}Density and dielectric constant at 1\,kHz of \sto  samples according to the shaping process.}
\begin{ruledtabular}
\begin{tabular}{p{0.3\linewidth} c c c} 
\textit{Shaping process} & \textit{Sample} & \textit{Density $d$} & $\epsr$/$\tgd$\,(1\,kHz)\\
 \hline
Tape casting \& & TC-STO-6a	& 67\% & 236/0.022 \\
 conventional & TC-STO-PM1	& 71\% & 253/0.02\\
 sintering  & TC-STO-1	& 73\% & ---\\
\hline
Uniaxial pressing	& UP-STO-2	&91\%  & 280/0.02\\
 \& conventional  & UP-STO-4	& 93\%&  280/0.015\\
 sintering  & UP-STO-7	& 95\% & 289/0.002\\
\hline
Spark & SPS-035	& $\!\geq\!99\%$ & 338/0.002\\
Plasma & SPS-036	& $\!\geq\!99\%$ &360/0.005 \\
Sintering & SPS-100	& $\!\geq\!99\%$ & 327/0.002\\
& SPS-101 & $\!\geq\!99\%$ & ---\\
\end{tabular}
\end{ruledtabular}
\end{table}

The dielectric constants were firstly compared at 1 kHz (Table\,\ref{tab:table1}). 
The relative permittivity of the TC samples slightly increases with the density $d$ from $\epsr =$ 236 to 253, while  
the dielectric losses remain around $\tgd = 0.02$. 
The relative permittivity of the UP increases from $\epsr =$ 280 to 289, as the density increases from $d\!=\!90$\% to 95\%, while the dielectric losses significantly decrease from $\tgd$ = 0.02 to 0.002. At last, the SPS samples exhibit the highest relative permittivity: $\epsr\!>\!320$. 
This process therefore yields very dense samples ($d\!\geq\!99\%$), whose relative permittivity at 1 kHz is the highest  ($\epsr\!>\!320$) of the three batches and whose dielectric losses are very low ($ \tgd\!\leq\!0.005$).

\begin{figure}[!htp]
\includegraphics[width =1\linewidth]{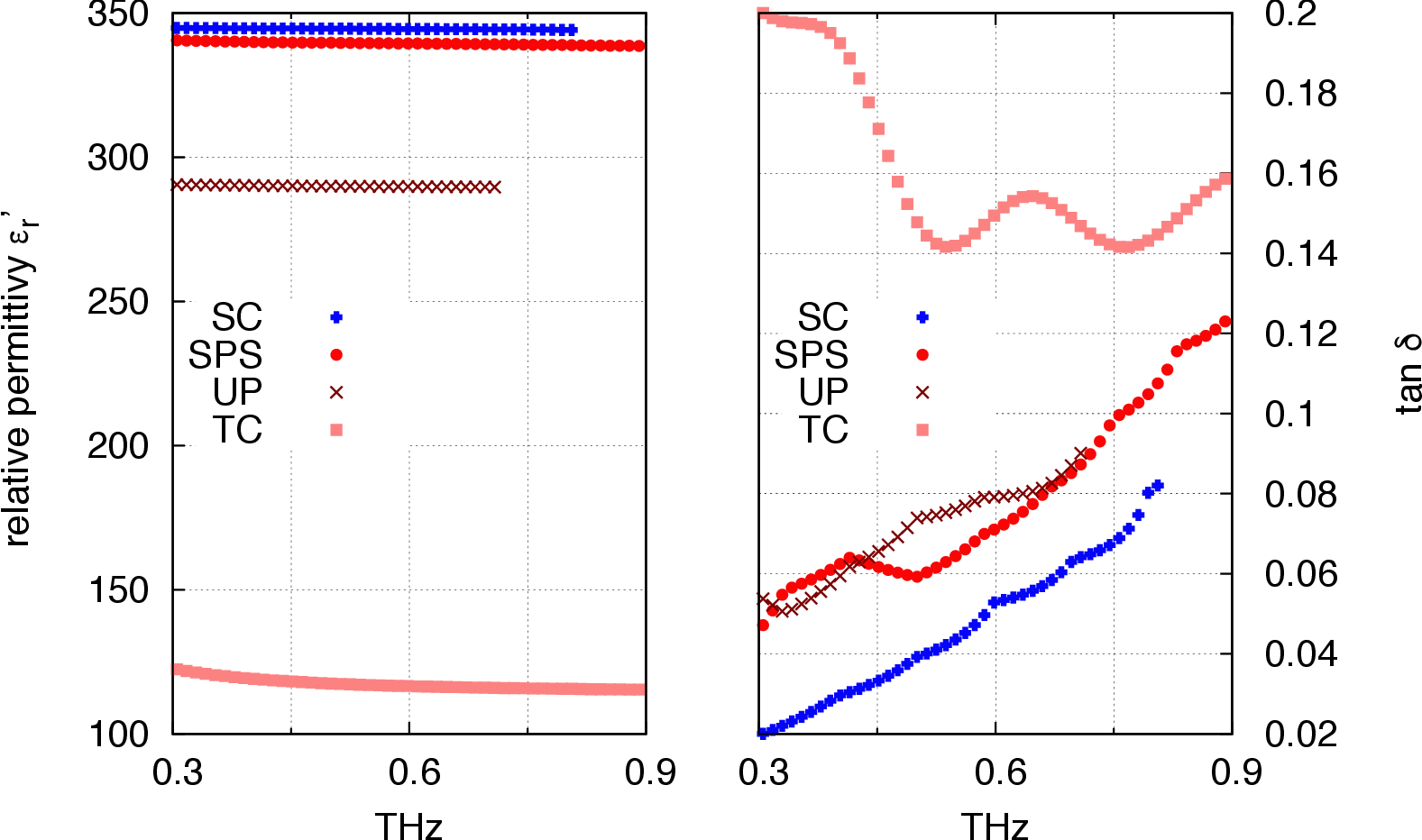}
\caption{\label{fig:relatperm} : Permittivity $\epsr$ and dielectric losses $\tgd$ in the terahertz range measured by THz-TDS of different processed \sto samples.}
\end{figure}

We further carried out the dielectric characterization in the terahertz range. 
A \sto single crystal \footnote{Verneuil growth and cubic perovskite structure -- CrysTec GmbH \url{http://www.crystec.de/}} serves as a reference in the THz set-up. Its dielectric constant was measured in the 0.2--0.9 THz range: the relative permittivity is $\epsr\!\simeq\!345$, while the dielectric losses practically linearly increase from $\tgd\!=\!0.02$ to 0.08 (Fig.\,\ref{fig:relatperm}).  These results are in good agreement with those of ref.\,\onlinecite{jjap48_matsumoto}.
%
The dielectric constant of three differently processed samples are reported in Fig.\,\ref{fig:relatperm} as well. The TC samples show low relative permittivity ($\epsilon_r\!\simeq\!117$) and high losses ($\tgd\!>\!0.14$). Due to the  porosity, the measured sample is a composite medium made of \sto and air, and its dielectric constant is an effective quantity $\epsf$ which is commonly described by the Bruggeman effective medium approximation (EMA).\cite{apl90_han} Porosity consequently lowers the relative permittivity $\epsr$ and increases the losses $\tgd$. 
The UP sample exhibits improved relative permittivity ($\epsr\!\simeq\!290$) and losses which also increase with the frequency in accordance with the PH model. 
The higher density similarly leads to increased relative permittivity $\epsr$ in the terahertz range.
Indeed, the relative permittivity of the SPS sample is actually close to that of the single crystal, i.e. $\epsr\!\simeq\!340$, while the dielectric losses, increasing from $\tgd\!\simeq\!0.055$ at 0.3 THz to $\simeq\!0.012$ at 0.9 THz, are about 1.3 times those of the single crystal. We ascribe this point to the coexistence of  Ti$^{4+}$ and Ti$^{3+}$ cations in the sample \footnote{A longer annealing after sintering could have lowered the numbers of Ti$^{3+}$ cations.} and to grain boundaries which do not exist in single crystals. Defects and grain boundaries may increase the mobility of atomic species and then losses. 
Fig.\,\ref{fig:relatperm} clearly demonstrates that the higher the density, the higher the relative permittivity $\epsr$ and the lower the dielectric losses $\tgd$.

\begin{figure}[!htp]
\includegraphics[width =\linewidth]{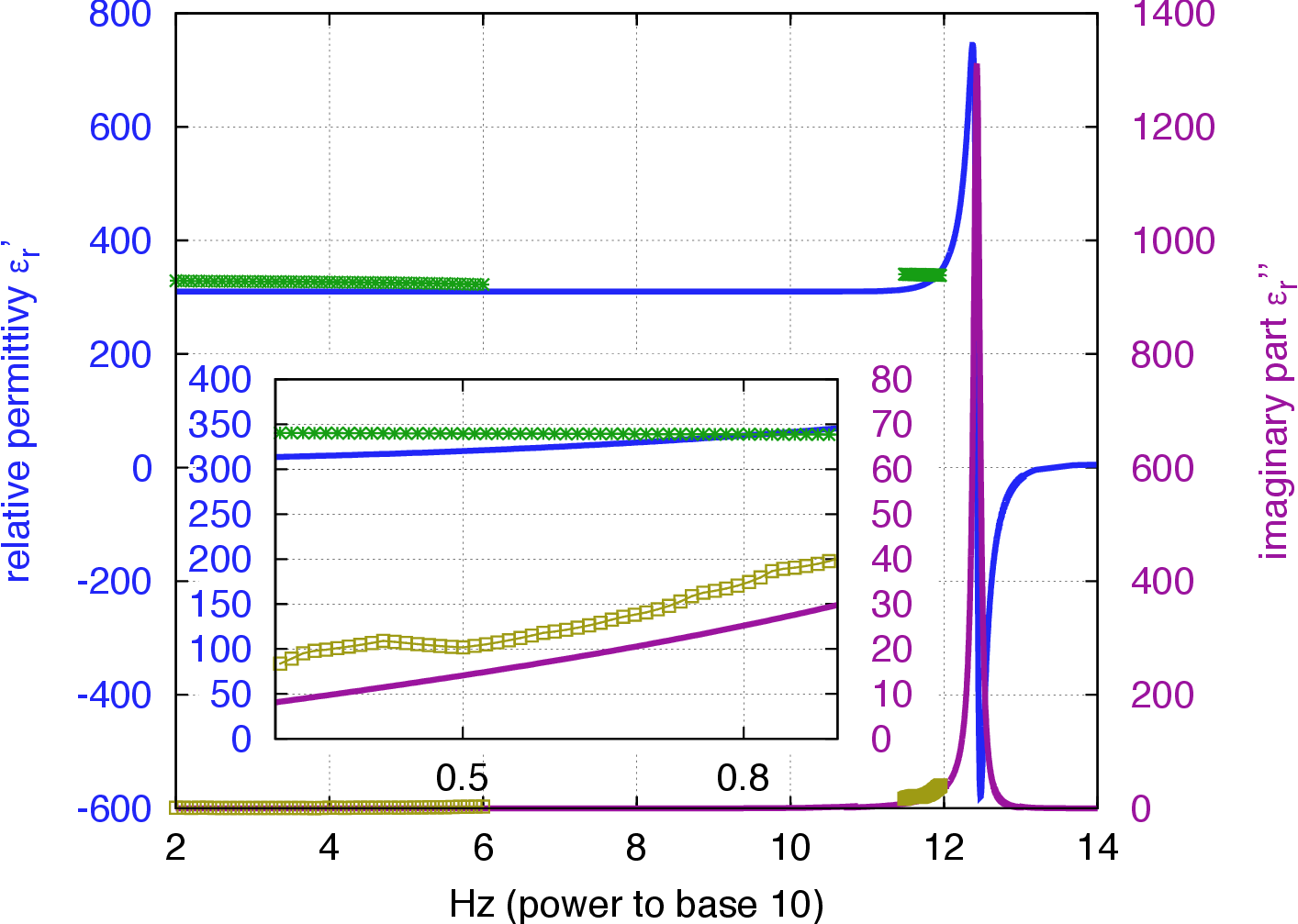}
\caption{\label{fig:perm_large_bande} : PH model of the dielectric response of \sto between $10^2$ and $10^{14}$\,Hz: $\epsrom = \epsr + \imath\,\epsilon_r''$ ($\epsilon_r'' = \tgd \times \epsr$)  and measurements in the low frequency (100Hz--1MHz) and the THz frequency ranges of SPS sample (green and yellow-green points). Inset : zoom between 0.3 and 0.9\,THz}
\end{figure}

The dielectric response $\epsrom$ of \sto is dispersive, because of the lattice vibrations, namely, the optical phonons.\cite{pr126_spitzer, pr145_barker} Their frequency is in the THz range, and we are concerned by
the Transverse Optical phonon of lowest frequency (TO1). The TO1 phonon frequency of \sto is 2.70THz.\cite{pr126_spitzer, pr145_barker} The dielectric function $\epsrom$ is described by the PH model (see eq.2 in ref.\,\onlinecite{apl90_han}), which is reported in fig.\ref{fig:perm_large_bande}. 
Losses (imaginary part of $\epsrom = \epsilon_r''$) resulting from the TO1 phonon greatly increase at terahertz frequencies. Measurements in both frequency ranges  are also reported in fig.\,\ref{fig:perm_large_bande} and are in good agreement with the PH model. 


Fully dense \sto material sintered by SPS, with a permittivity as high as that of the single crystal, is suitable for ADM applications at THz frequencies. The dielectric constant is in good agreement with the PH model. SPS is the effective process required by ADMs to develop. 


This work was supported by the Agence Nationale de la Recherche under the TŽraMŽtaDiel convention (grant ANR-12-BS03-0009). \'E.A. thanks Fabrice Rossignol for his help. 


\bibliography{sto_terametadiel_20171128} 

\end{document}